\documentclass[twocolumn,showpacs,preprintnumbers,amsmath,amssymb]{revtex4}

\usepackage{subfigure}
\usepackage{graphicx}
\usepackage{dcolumn}
\usepackage{amsfonts,amsmath,amssymb,bm}
\begin{document}
\title{Persistent current and Drude weight for the one-dimensional Hubbard model from current lattice density functional theory}

\author{A. Akande and S. Sanvito \thanks{akandea@tcd.ie}}
\affiliation{School of Physics and CRANN, Trinity College, Dublin 2, Ireland}

\date{\today}

\begin{abstract}
The Bethe-Ansatz local density approximation (LDA) to lattice density functional theory (LDFT) for the one-dimensional repulsive Hubbard model 
is extended to current-LDFT (CLDFT). The transport properties of mesoscopic Hubbard rings threaded by a magnetic flux are then systematically 
investigated by this scheme. In particular we present calculations of ground state energies, persistent currents and Drude weights for both a repulsive 
homogeneous and a single impurity Hubbard model. Our results for the ground state energies in the metallic phase compares favorably well 
with those obtained with numerically accurate many-body techniques. Also the dependence of the persistent currents on the Coulomb and the impurity 
interaction strength, and on the ring size are all well captured by LDA-CLDFT. Our study demonstrates that CLDFT is a powerful tool for studying 
one-dimensional correlated electron systems with high accuracy and low computational costs. 
\end{abstract}


\maketitle
\section{Introduction}


Quantum dots, routinely made by electrostatically confining a two-dimensional electron gas \cite{Davies}, have been extensively studied in recent 
years \cite{Thierry_book}. The interest in these low-dimensional structures stems from the fact that their physics is controlled by quantum effects. 
Furthermore, while sharing many similarities with real atoms, quantum dots manifest intriguing low-energy quantum phenomena, which are specific to them. 
This is because their properties can be influenced by external factors such as the geometry or the shape of the confining potential and the application 
of external fields. Clearly some of these features are not accessible in real atoms. Research in the past has been motivated by the possibility of developing 
novel quantum dot based devices in both the fields of quantum cryptography/computing \cite{Zipper} and spintronic \cite{Foldi}, as well as by the simple 
curiosity of exploring the properties of many-electron systems in reduced dimensions. 


Quantum rings represent a particular class of quantum dots \cite{Lorke, Fuhrer}, where electrons are confined in circular regions \cite{Garcia, Lorke2}.
The circular geometry can sustain an electrical current, which in turns can be induced by threading a magnetic flux across the ring itself. Such a 
magnetic flux produces exciting effects like Ahanorov-Bohm (AB) oscillations \cite{Aharonov, Levy} and persistent currents \cite{Buttiker}, 
effects that were anticipated as early as the late 60's \cite{Bloch1, Bloch2}. In one dimension (1D) the persistent currents have been thoroughly 
studied \cite{Buttiker, Buttiker2}. These, as many other physical properties of the ring, are a periodic function of the magnetic flux quantum, 
$\Phi_0=hc/e$ ($h$ is the Planck's constant, $c$ the speed of light and $e$ the electron's charge). 


A number of earlier theoretical studies \cite{Giancarlo, Kirchner, Molina, SKMaiti2004, Thierry2} on persistent currents focused on unveiling
the role of electron correlations and disorder over the electron transport. This line of research is inspired by the fact that electronic correlation in 1D 
always leads to non-fermionic low-energy quasiparticle excitations. In fact, even in the presence of weak interaction, 1D fermions 
behave differently from a Fermi liquid and their ground state is generally referred to as Luttinger liquid. This possesses specific collective 
excitations \cite{Kolomeisky}.


There are two theoretical frameworks commonly used to study finite 1D rings \cite{Viefers}. The first is based on the continuum model, where electrons
move in a uniform neutralizing positive background and interact via Coulomb repulsion, $e^2/4\pi\varepsilon_0r$ ($\varepsilon_0$ is the 
vacuum permittivity, $e$ the electron charge, $r$ the distance between two electrons). The second is populated by lattice models, where the electronic 
structure is written in a tight-binding form and the electron-electron 
interaction is commonly described at the level of Hubbard Hamiltonian \cite{Hubbard1, Hubbard2}. In both frameworks exact diagonalization (ED) 
has been the preferential solving strategy for small systems (small number of sites and electrons) \cite{SKMaiti2004,Bouzerar}. Additional methods 
used to study quantum rings over lattice models include Bethe Ansatz (BA) \cite{Bethe, LiebWu}, renormalization group \cite{Meden} and 
density matrix renormalization group \cite{Dias}. In contrast the continuum model has been tackled with self-consistent Hartree Fock 
techniques \cite{Cohen}, bosonization schemes \cite{Gogolin}, conformal field theory \cite{Jaimungal}, current-spin density functional theory 
\cite{Viefer2} and quantum Monte Carlo \cite{Emperador}.


Many of the methods developed for solving lattice models for interacting electrons suffer from a number of intrinsic limitations connected either to
their large computational overheads or to the need of using a drastically contracted Hilbert space. Density functional theory (DFT) can be a natural 
solution to these limitations. DFT is a highly efficient and precisely formulated method \cite{HK, KS}, originally developed for the Coulomb 
interaction (this is commonly known as {\it ab initio} DFT), and then extended to lattice models \cite{Gunnarsson, Schonhammer, Capelle3}. 
Lattice DFT (LDFT) is based on the rigorously proved statement that the ground state of an interacting electron system is a universal functional 
of the local site occupation. The functional, as in {\it ab initio} DFT, is unknown explicitly. However all the many-body contributions to the total energy 
can be incorporated in a single term, the exchange and correlation (XC) energy, for which a hierarchy of approximations can be constructed.


The most commonly used approximation for the XC energy in {\it ab initio} DFT is probably the local density approximation (LDA) \cite{KS,Densityfunctional},
where the exact (unknown) XC energy is replaced by that of the homogeneous electron gas. The theory is then expected to work best in situations
close to those described by the {\it reference} system, i.e. close to the homogeneous electron gas. Since in 1D Fermi liquid theory breaks down, the homogeneous 
electron gas is no longer a good reference. For the homogeneous Hubbard model it was then proposed \cite{Capelle3, Capelle1} to use instead the BA 
construction of Lieb and Wu \cite{LiebWu}. Such a scheme was then applied successfully to a wide range of situations \cite{Capelle2, Silva, Campo, Xianlong1, 
Xianlong2, Xianlong3,Akin} and more recently it has been extended to time-dependent problems \cite{Xianlong3, Verdozzi, Kurth,Stefanucci} 
and to the 3D Hubbard model \cite{Karlsson}. 

LDFT can be further extended to include the action of a vector potential,  i.e. it can be used to tackle problems where a magnetic flux is relevant. This 
effectively corresponds to the construction of current-LDFT (CLDFT). Such an extension of LDFT was proposed recently for one-dimensional spinless 
fermions with nearest-neighbor interaction \cite{Schenk} and it is here adapted to the repulsive Hubbard model. The newly constructed functional is then 
used to investigate total energies, persistent currents and Drude weights of a mesoscopic repulsive Hubbard ring threaded by a magnetic flux. 

The paper is organized as follows. Section II reviews the theoretical foundations leading to the construction of CLDFT and to its LDA. Then we present 
our results for both homogeneous and defective rings, highlighting the main capabilities and limitations of our scheme, and finally we conclude. 

\section{Theoretical Formulation of Current Lattice DFT}

Current-DFT (CDFT) is a generalization of time-dependent density functional theory \cite{Gross} to include in the Hamiltonian an external vector potential 
\cite{thebook}. In this case the theory is constructed over two fundamental quantities, namely the electron density, $n$, and the paramagnetic 
current density, $\vec{j}_p$. The Hohenberg-Kohn theorem \cite{HK} is thus expanded to the statement that the ground state $n$ and 
$\vec{j}_p$ uniquely determine the ground-state wave-function and consequently the expectation values of all the operators \cite{Vignale1, Vignale2}. 
Equally important is the fact that the standard Kohn-Sham construction can also be employed for CDFT, so that the many-body problem can be 
mapped onto a fictitious single-particle one, with the two sharing the same ground state $n$ and $\vec{j}_p$ \cite{Vignale1, Vignale2}. Practically 
one then needs to solve self-consistently a system of single-particle equations. Also for CDFT all the unknown of the theory are incorporated in the 
XC energy, which then needs to be approximated. 

The scope of this section is to describe how {\it ab initio} CDFT has been translated to lattice models and how a suitable approximation for the 
XC energy associated to the Hubbard Hamiltonian can be constructed. Our description follows closely the one previously given by Dzierzawa 
et al. \cite{Schenk}. In general a vector potential, $\vec{A}$, enters into a lattice model via Peierls substitution \cite{Peir,Vogel}, where the matrix 
elements of the $\vec{A}$-dependent Hamiltonian, $H(\vec{r},\vec{p}+\frac{e}{c}\vec{A})$, can be written in terms of those for $\vec{A}=0$ as
\begin{equation}
\langle\vec{R}^\prime|H(\vec{r},\vec{p}+\frac{e}{c}\vec{A})|\vec{R}\rangle=
\langle\vec{R}^\prime|H(\vec{r},\vec{p})|\vec{R}\rangle 
\mathrm{e}^{-\frac{ie}{\hbar c}\int_{\vec{R}}^{\vec{R}^\prime}\vec{A}\cdot\mathrm{d}\vec{s}}\:,
\end{equation}
where $c$ is the speed of light and $|\vec{R}\rangle$ is the generic orbital located at the position $\vec{R}$ and belonging to the basis set (here 
assumed orthogonal) used to construct the tight-binding Hamiltonian. 

When the Peierls substitution is applied to the construction leading to the 1D Hubbard model the only term in the Hamiltonian that gets 
modified is the kinetic energy $\hat{T}$. This takes the form
\begin{equation}
\hat{T}=-t \sum_{\sigma,\:l=1}^{L}(e^{-i{\Phi_{\sigma l}/L}}\:\hat{c}^\dagger_{\sigma\:l+1}\hat{c}_{\sigma\:l}+hc)\:,
\label{Ekin}
\end{equation}
where we have considered a system comprising $L$ atomic sites (note that the ring boundary conditions imply $L+1=1$). 
In the equation (\ref{Ekin}) $\hat{c}^\dagger_{\sigma l}$ ($\hat{c}_{\sigma l}$) is the creation (annihilation) operator for an electron 
of spin $\sigma$ ($\sigma=\uparrow, \downarrow$) at the $l$-site, $t$ is the hopping integral and $\Phi_{\sigma l}$ is the phase 
associated to the $l$-th bond, which effectively describes the action of $\vec{A}$. The remaining terms in the Hamiltonian 
are unchanged so that the 1D Hubbard model in the presence of a vector potential is defined by
\begin{equation}\label{Hint}
\hat{H}^\Phi_\mathrm{Hubbard}=\hat{T}+\hat{U}+\sum_l^Lv_l^\mathrm{ext}\hat{n}_{l}\:, 
\end{equation}
where \{$v_l^\mathrm{ext}$\} is the external potential ($v_l^\mathrm{ext}$ is the on-site energy of the $l$-site), while the Coulomb 
repulsion term is $\hat{U}=U\sum_{l=1}^L \hat{n}_{\uparrow l} \hat{n}_{\downarrow l}$, with $U$ being the Coulomb repulsion energy
and $\hat{n}_{\sigma l}=\hat{c}^\dagger_{\sigma l}\hat{c}_{\sigma l}$. Throughout this work we always consider the diamagnetic 
(non-spin polarized) case so that $\Phi_{\uparrow l}=\Phi_{\downarrow l}=\Phi_l$ and $n_{\uparrow l}=n_{\downarrow l}=n_l$. 

The first step in the construction of a CLDFT is the formulation of the problem in a functional form.
The basic variables of the theory are the site occupation $n_l=\langle\Psi|\hat{n}_l|\Psi\rangle$ and bond paramagnetic current,  
$j_l=\langle\Psi|\hat{j}_l|\Psi\rangle$, where $|\Psi\rangle$ is the many-body wavefunction and the bond paramagnetic current 
operator is defined as
\begin{equation}\label{paraJ}
\hat{j}_l=-it(e^{-i{\Phi_l/L}}\:\hat{c}^\dagger_{\sigma\:l+1}\hat{c}_{\sigma\:l}-hc)
\end{equation}
In complete analogy to {\it ab initio} CDFT we can write the total energy, $\mathcal{E}$, of the Hamiltonian (\ref{Hint}) as a functional 
of the local external potentials and phases
\begin{equation}\label{LF1}
\mathcal{E}=\mathcal{F}[n_l,j_l]+\sum_lv_l^\mathrm{ext}n_l+\sum_l\Phi_lj_l, 
\end{equation}
so that
\begin{equation}
\begin{aligned}\label{LF2}
n_l=\langle\hat{n}_l\rangle&=\frac{\partial \mathcal{E}}{\partial v_l^\mathrm{ext}}\:,\\
j_l=\langle\hat{j}_l\rangle&=\frac{\partial \mathcal{E}}{\partial \Phi_l}\:.
\end{aligned}
\end{equation}
$\mathcal{F}[n_l,j_l]$ is a universal functional, whose functional derivatives with respect to $\{n_l\}$ and $\{j_l\}$ satisfy the following two equations
\begin{equation}\label{LF3}
\begin{aligned}
v_l^\mathrm{ext}=&-\frac{\partial \mathcal{F}}{\partial n_l}\\
\Phi_l=&-\frac{\partial \mathcal{F}}{\partial j_l}.
\end{aligned}
\end{equation}
Note that equations (\ref{LF1}) through (\ref{LF3}) follow directly from the properties of the Legendre transformation. 

In order to make the theory practical one has now to introduce the auxiliary single-particle Kohn-Sham system. This is described by a single-particle 
Hamiltonian, $\hat{H}^\mathrm{s}$, whose ground state site occupations and bond paramagnetic currents are identical to those of the interacting 
system [described by equation (\ref{Hint})]. $\hat{H}^\mathrm{s}$ reads
\begin{equation}\label{Hs}
\hat{H}^\mathrm{s}=\hat{T}^\mathrm{s}+\sum_l^Lv_l^\mathrm{s}\hat{n}_{l}, 
\end{equation}
where $\hat{T}^\mathrm{s}=-t \sum_{\sigma,l=1}^{L-1}(e^{-i{\Phi_l^\mathrm{s}/L}}\:\hat{c}^\dagger_{\sigma\:l+1}\hat{c}_{\sigma l}+hc)$ and the
associated local effective potentials and phases are $v_l^\mathrm{s}$ and $\Phi_l^\mathrm{s}$ respectively. The single-particle Schr\"{o}dinger 
equation is then
\begin{equation}
 \hat{H}^\mathrm{s}|\Psi^\mathrm{s}_\alpha\rangle=\epsilon_\alpha|\Psi^\mathrm{s}_\alpha\rangle,
\end{equation}
and the site occupation is defined as
\begin{equation}\label{occup}
n_l^\mathrm{s}=\sum_\alpha f_\alpha\langle\Psi^\mathrm{s}_\alpha|\hat{n}_l|\Psi^\mathrm{s}_\alpha\rangle\:,
\end{equation}
where $f_\alpha$ is the occupation number. An analogous expression can be written for $j^\mathrm{s}_l$. 

The energy functional associated the Kohn-Sham system, $\mathcal{F}^\mathrm{s}$, can be constructed by performing again a Legendre 
transformation
\begin{equation}\label{fs}
\mathcal{F}^\mathrm{s}=\mathcal{E}^\mathrm{s}-\sum_lv_l^\mathrm{s}n_l^\mathrm{s}-\sum_l\Phi_l^\mathrm{s}j_l^\mathrm{s}\:,
\end{equation}
where $\mathcal{E}^\mathrm{s}$ is the total energy of the single-particle system and the following two equations are valid
\begin{equation}\label{LF4}
\begin{aligned}
v_l^\mathrm{s}=&-\frac{\partial \mathcal{F}^\mathrm{s}}{\partial n_l^\mathrm{s}}\:,\\
\Phi_l^\mathrm{s}=&-\frac{\partial \mathcal{F}^\mathrm{s}}{\partial j_l^\mathrm{s}}\:.
\end{aligned}
\end{equation}
The crucial point is that in the ground state the real and the Kohn-Sham systems share the same site occupation and paramagnetic
current, i.e. $n_l=n_l^\mathrm{s}$ and $j_l=j_l^\mathrm{s}$.

Thus one is now in the position of defining the XC energy, $\mathcal{E}^\mathrm{xc}$, as usual, i.e. as the difference between $\mathcal{F}$ 
for the interacting and the Kohn-Sham systems after the classical Hartree energy $\mathcal{E}^\mathrm{H}$ has also been subtracted,
\begin{equation}\label{xcene}
\mathcal{E}^\mathrm{xc}[n_l,j_l]=\mathcal{F}[n_l,j_l]-\mathcal{F}^\mathrm{s}[n_l,j_l]-\mathcal{E}^\mathrm{H}[n_l]\:. 
\end{equation}
Note that for all the functionals in equation (\ref{xcene}) we took the short notation $\{n_l\}\rightarrow n_l$ and $\{j_l\}\rightarrow j_l$, i.e. 
the functionals depend on all the on-site occupations and bond paramagnetic currents. The single-particle effective potentials and phases
can now be defined. In fact by taking the functional derivative of equation (\ref{xcene}) with respect to $n_l$ and $j_l$ and by using 
the equations (\ref{LF3}) and (\ref{LF4}) one obtains
\begin{equation}\label{kspot}
 \begin{aligned}
  v_l^\mathrm{s}=&v_l^\mathrm{ext}+v_l^\mathrm{H}+v_l^\mathrm{xc}\:,\\
  \Phi_l^\mathrm{s}=&\Phi_l+\Phi_l^\mathrm{xc}\:,
 \end{aligned}
\end{equation}
where
\begin{equation}
 \begin{aligned}
  v_l^\mathrm{xc}=&\frac{\partial \mathcal{E}_l^\mathrm{xc}}{\partial n_l}\:,\\
  \Phi_l^\mathrm{xc}=&\frac{\partial \mathcal{E}_l^\mathrm{xc}}{\partial j_l}\:,
 \end{aligned}
\end{equation}
and $v_l^\mathrm{H}=\partial \mathcal{E}_l^\mathrm{H}/\partial n_l$ ($=Un_l/2$) is the Hartree potential.

Finally $\mathcal{E}^\mathrm{xc}$ can be re-written in terms of the expectation values of the original Hamiltonian. In fact by substituting
the functional forms of $\mathcal{F}$ and $\mathcal{F}^\mathrm{s}$ into the equation (\ref{xcene}),
\begin{equation}\label{exc2}
\mathcal{E}^\mathrm{xc}=\mathcal{E}-\mathcal{E}^\mathrm{s}+\sum_l(v_l^\mathrm{s}-v_l^\mathrm{ext})n_l+\sum_l
(\Phi_l^\mathrm{s}-\Phi_l)j_l-\mathcal{E}^\mathrm{H}[n_l]\:,
\end{equation}
by using the equations (\ref{Hint}) and (\ref{Hs}),
\begin{equation}\label{vsvl}
\sum_l(v_l^\mathrm{s}-v_l^\mathrm{ext})n_l=\mathcal{E}^\mathrm{s}-\mathcal{E}-\langle\Psi^\mathrm{s}|\hat{T}^\mathrm{s}|\Psi^\mathrm{s}
\rangle+ \langle\Psi|\hat{T}+\hat{U}|\Psi\rangle\:,
\end{equation}
and again by substituting equation (\ref{vsvl}) into equation (\ref{exc2}), one obtains a close expression for the XC energy
\begin{equation}\label{exc3}
\mathcal{E}^\mathrm{xc}=\langle\Psi|\hat{T}+\hat{U}|\Psi\rangle-\langle\Psi^\mathrm{s}|\hat{T}^\mathrm{s}|\Psi^\mathrm{s}
\rangle+\sum_l(\Phi_l^\mathrm{s}-\Phi_l)j_l-\mathcal{E}^\mathrm{H}[n_l]. 
\end{equation}
 
Once the theory is formally established the remaining task is that of finding an appropriate approximation for
$\mathcal{E}^\mathrm{xc}$. As for the case of standard LDFT \cite{Capelle1,Capelle2}, the strategy here is that of considering 
the BA solution for the homogeneous limit of $\hat{H}^\Phi_\mathrm{Hubbard}$ (this is defined in equation (\ref{Hint}) by setting 
$v_l=v$ and $\Phi_l=\Phi$) and then of taking its local density approximation $n\rightarrow n_l$, $\Phi\rightarrow\Phi_l$ \cite{Schenk}, i.e. 
\begin{equation}
\mathcal{E}^\mathrm{xc}_\mathrm{LDA}[n_l,j_l]=\sum_le^\mathrm{xc}[n_l,j_l], 
\end{equation}
where $e^\mathrm{xc}[n,j]={\mathcal{E}^\mathrm{xc}[n,j]/L}$ is the XC energy density (per site) of the homogeneous system. The first term of the 
equation (\ref{exc3}) can be calculated exactly using the BA procedure \cite{Shastry}. This provides the ground state energy as a function 
of $n$ and $\Phi$, so that one still needs to re-express it in terms of $n$ and $j$. However the phase variable $\Phi$ can be eliminated from 
the ground state energy by using 
\begin{equation}
j=\frac{\partial \mathcal{E}(n,\Phi)}{\partial \Phi}\:. 
\end{equation}

Thus finally one can explicitly write $e^\mathrm{xc}(n,j)$ (the full derivation for the 1D Hubbard Hamiltonian is presented in 
the Appendix)
\begin{equation}\label{exc51}
e^\mathrm{xc}(n,j)=e^\mathrm{xc}(n,0)+\frac{1}{2}\Lambda^\mathrm{xc}(n)j^2,
\end{equation}
where
\begin{equation}
\begin{aligned}
e^\mathrm{xc}(n,0)=&\frac{\mathcal{E}^\mathrm{BA}(n,0)-\mathcal{E}^0(n,0)-\mathcal{E}^\mathrm{H}(n)}{L}\:,\\
\Lambda^\mathrm{xc}(n)=&\frac{1}{2}\left [\frac{1}{D_c^0(n)}-\frac{1}{D_c^\mathrm{BA}(n)}\right ].
\end{aligned}
\end{equation}
In the equations above $\mathcal{E}^0(n,0)$ and $D_c^0(n)$ are respectively the non-interacting ground state energy and charge stiffness, 
while $\mathcal{E}^\mathrm{BA}(n,0)$ and $D_c^\mathrm{BA}(n)$ are the same quantities for the interacting case as calculated from the BA. 
Finally, the XC contributions to the Kohn-Sham potential can be obtained by simple functional derivative (in this case by simple derivative)
of the exchange and correlation energy density with respect to the fundamental variables $n$ and $j$, i.e.
\begin{equation}\label{vxc1}
v^\mathrm{xc}_\mathrm{BALDA}(n_l,j_l)=\left.\frac{\partial e^\mathrm{xc}(n,j)}{\partial n}\right|_{n\rightarrow n_l,j\rightarrow j_l}\:,
\end{equation}
and
\begin{equation}\label{phixc1}
\Phi^\mathrm{xc}_\mathrm{BALDA}(n_l,j_l)=\left.\frac{\partial e^\mathrm{xc}(n,j)}{\partial j}=\Lambda^\mathrm{xc}(n)j\right|_{n\rightarrow n_l,j\rightarrow j_l}\:,
\end{equation}
where BALDA, as usual, stands for Bethe Ansatz local density approximation.

\begin{figure}[htb]
\begin{center}
\includegraphics[width=0.48\textwidth,angle=0.0]{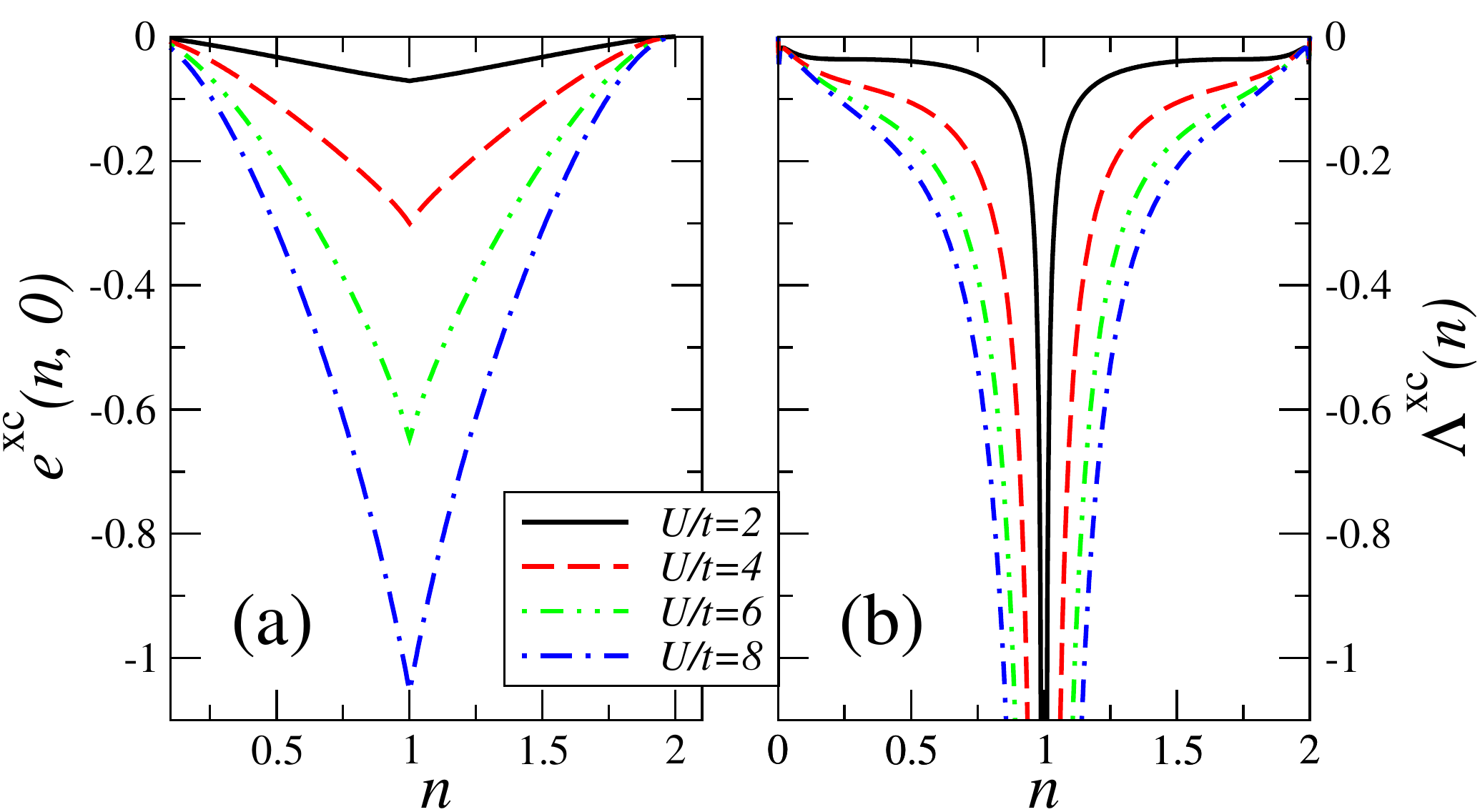}
\end{center}
\caption {(Color online) The XC energy density (per site) for a homogeneous 1D Hubbard ring threaded by a magnetic flux as a function of the 
electron density and for different values of interaction strength $U/t$: (a) $e^{xc}(n,0)$ and (b) $\Lambda^{xc}(n)$.} \label{exchange_lambda}
\end{figure}
In the two panels of figure \ref{exchange_lambda} we present $e^\mathrm{xc}(n,0)$ and $\Lambda^\mathrm{xc}(n)$ as a function of the electron 
density, $n$, for different interaction strengths $U/t$. As in the case of standard LDFT also for CLDFT there is a divergence in the $n$-derivative of both 
$e^\mathrm{xc}(n,0)$ and $\Lambda^\mathrm{xc}(n)$ at half-filing ($n=1$). This is in correspondence of the metal-insulator-transition present in the 
1D Hubbard model for finite $U/t$. In the case of $\Lambda^\mathrm{xc}(n)$ the divergence is also in $\Lambda^\mathrm{xc}(n)$ itself.

The solution of the Kohn-Sham problem proceeds as follows. First an initial guess for the site occupations is used to construct the initial local 
paramagnetic current density. Then, the functional derivatives of equations (\ref{vxc1}) and (\ref{phixc1}) are evaluated at these given $n$ and 
$j$ so that the Kohn-Sham potential is constructed. The Kohn-Sham equations are then solved to obtain the new set of Kohn-Sham orbitals 
from which the new orbital occupations and  bond paramagnetic currents are calculated [by using equation (\ref{occup})]. The procedure is then 
repeated untill self-consistency is reached, i.e. until the potentials (or the densities) at two consecutive iterations vary below a certain threshold. 
After convergence is achieved the total energy for the interacting system is calculated from 
\begin{equation}\label{totalenergy2}
 \mathcal{E}=\sum_\alpha f_\alpha\epsilon_\alpha+\mathcal{E}^\mathrm{xc}[n_l,j_l]-\mathcal{E}^\mathrm{H}[n_l]-\sum_lv_l^\mathrm{xc}n_l,
\end{equation}
where the first term is the sum of single-particle energies and the other terms are the so-called \textit{double counting} corrections.

\section{Results and Discussion}

We now discuss how CLDFT performs in describing both the energetics and the transport properties of 1D Hubbard rings in presence of a magnetic
flux. For small rings our results will be compared with those obtained by diagonalizing exactly the Hamiltonian of equation (\ref{Hint}), while 
CLDFT for large rings will be compared with the BA solution. First we will consider homogeneous rings and then we will explore the single impurity
problem.

\subsection{Homogeneous rings: general properties}

In this section we focus our attention on discussing the general features of CLDFT applied to homogeneous Hubbard rings threatened by
a magnetic flux, i.e. on the performance of CLDFT in describing the Ahanorov-Bohm effect. We start our analysis by comparing the CLDFT
results with those obtained by ED. Since ED is numerically intensive such a comparison is limited to small systems. 

\begin{figure}[htb]
\begin{center}
\includegraphics[width=0.48\textwidth,angle=0.0]{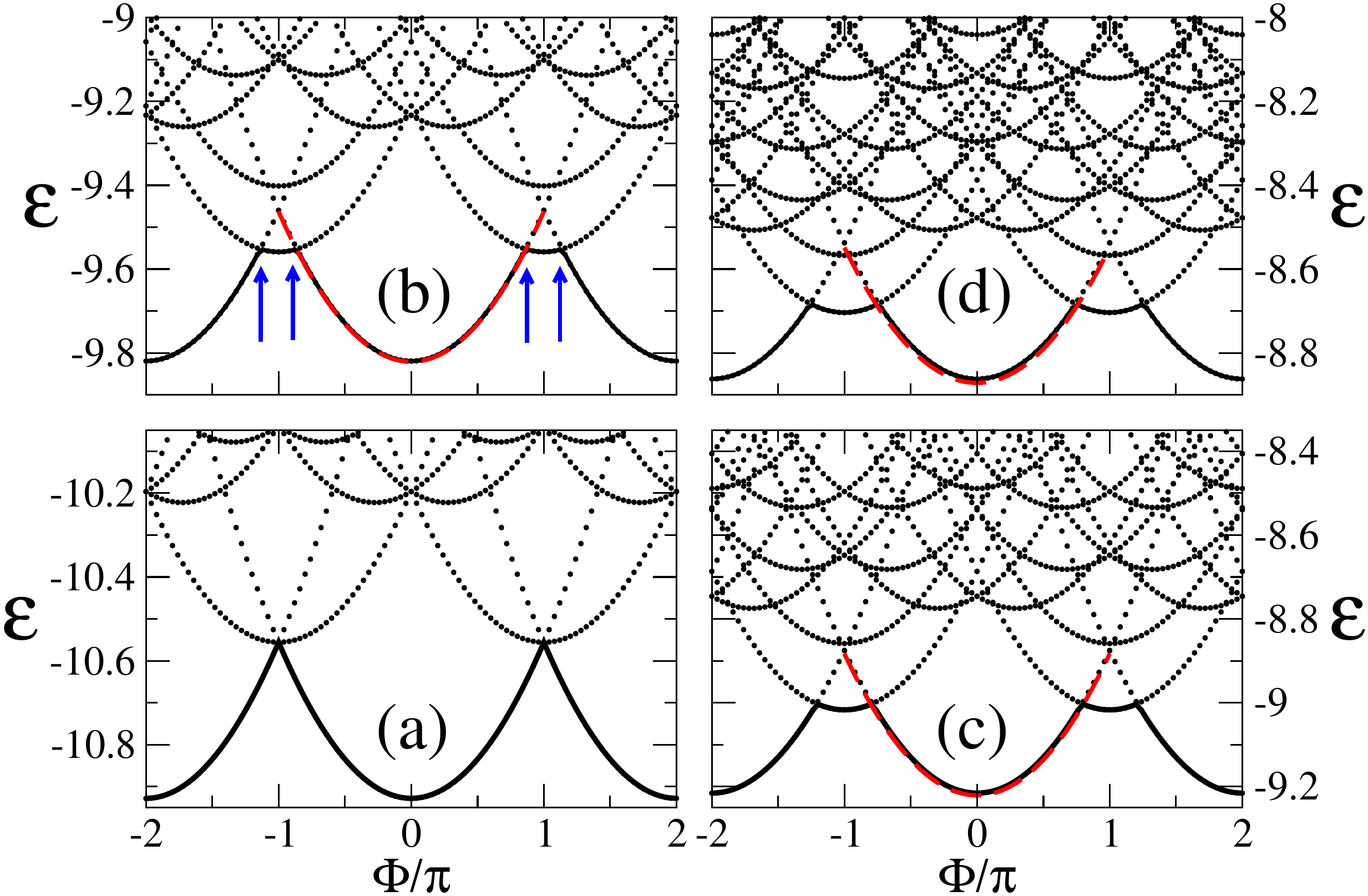}
\end{center}
\caption {(Color online) The low lying energy spectrum, $\mathcal{E}$, of a 12-site ring at quarter filling ($n=1/2$) as a function of the magnetic 
flux, $\Phi$, and calculated for different interaction strengths $U/t$. The black dotted lines represent ED results while the dashed red ones are 
for CLDFT. Note that for the non interacting case, $U/t=0$, in panel (a) there is no difference between CLDFT and ED. Panels (b)-(d) are for
the interacting case at different interaction strengths: (b) $U/t=2$, (c) $U/t=4$ and (d) $U/t=6$. In panel (b) the blue arrows indicate the region 
where the triplet state becomes the ground state.} \label{12_6low_lying_stateU0_U2_U4_U6}
\end{figure}

In figure \ref{12_6low_lying_stateU0_U2_U4_U6} we present the first low-lying energy levels, $\mathcal{E}$, calculated by ED as a function 
of the magnetic flux, $\Phi$, for a small 12-site ring at quarter filling ($n=1/2$). In particular we present results for the non-interacting case 
[panel (a)] and for the interacting one at three different interaction strengths: (b) $U/t=2$, (c) $U/t=4$ and (d) $U/t=6$. Exact results (ED) are
in black, while those obtained with CLDFT in red. In general the ground state energy is mimimized at $\Phi=0$ when the number of electrons 
is $N=4m+2$ and at $\Phi=\pi$ for $N=4m$, with $m$ being an integer \cite{Nakano}. Here we consider the case $N=4m+2$ where the ground 
state is a singlet \cite{Stafford}. 

For non-interacting electrons, $U/t=0$, the total energy of the singlet ground state is a parabolic function of $\Phi$. Also the various 
excited states have a parabolic dependence on $\Phi$ and simply correspond to single-particle levels with different wave-vectors.
As the electron-electron interaction is turned on the non-interacting spectrum gets modified in two ways. Firstly there is a second branch 
in the ground state energy as a function of $\Phi$ appearing at around $\Phi=\pm\pi$ (see the blue arrows in panel (b) of 
Fig.~\ref{12_6low_lying_stateU0_U2_U4_U6}). This originates from the degeneracy lifting between the single and the triplet solution at 
$\Phi=\pm\pi$, with the triplet being pushed down in energy and becoming the ground state. The $\Phi$ region where the ground state 
is a triplet widens as the interaction strengths increases. The second effect is the expected reduction of the ground state total energy 
as a function of $U/t$.

Since CLDFT is a ground state theory, it provides access only to the ground state energy, $\mathcal{E}$. This is calculated next and plotted 
in figure \ref{12_6low_lying_stateU0_U2_U4_U6} in the interval $-\pi\le\Phi\le\pi$ for different $U/t$. As one can clearly see from the figure
the performance of CLDFT is rather remarkable, to a point that the CLDFT energy is practically identical to that calculated with ED. However
CLDFT completely misses the cusps in the $\mathcal{E}(\Phi)$ profile arising from the crossover between the singlet and the triplet state.
Level crossing invalidates the BA approximation leading to the interacting XC energy [see equation (\ref{fluxdependence3}) in the appendix]
and so failures are expected \cite{Romer}. This observation is in agreement with earlier studies \cite{Viefers} in which the inability of CDFT 
to reproduce level crossing was already noted. Nevertheless, as long as the singlet remains the ground state, the agreement between 
CLDFT and ED results is remarkable, even if this small ring is rather far from being a good approximation of the thermodynamic limit (the BA
solution) upon which the functional has been constructed.

\begin{figure}[htb]
\begin{center}
\includegraphics[width=0.43\textwidth,angle=0.0]{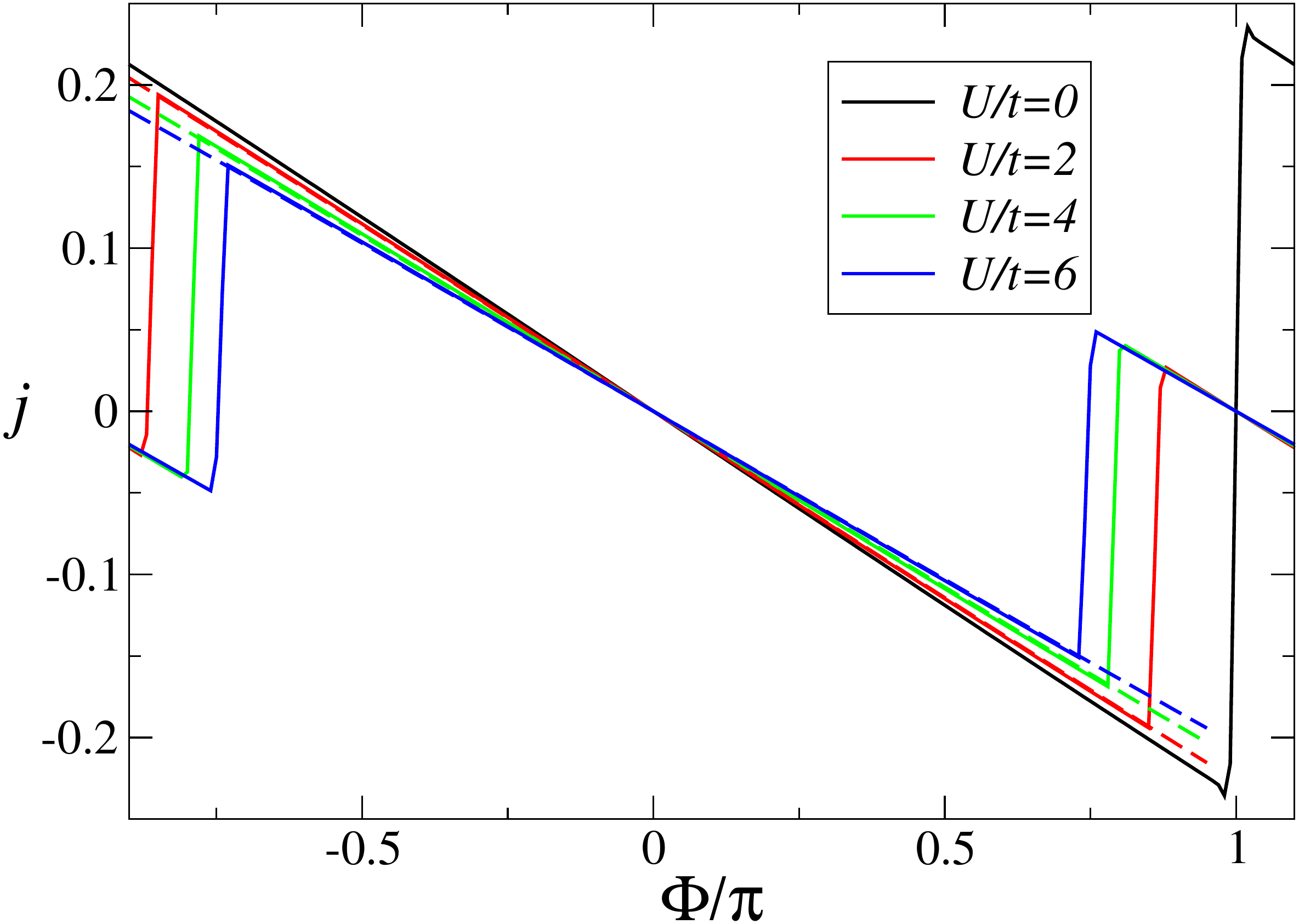}
\end{center}
\caption {(Color online) Persistent current profile, $j$, for a 12-site ring at quarter filling ($n=1/2$) obtained with both ED and CLDFT for 
different $U/t$. The full lines are the $j$ calculated with ED while the dashed ones are for CLDFT.} \label{12_6persistentU0_U2_U4_U6}
\end{figure}
Having calculated the total energies with both ED and CLDFT, the corresponding persistent currents, $j$, can be obtained by taking the 
numerical derivative of $\mathcal{E}(\Phi)$ with respect to $\Phi$. In figure \ref{12_6persistentU0_U2_U4_U6} we show results for the 
12 site ring at quarter filling ($n=1/2$), whose total energy was presented in figure \ref{12_6low_lying_stateU0_U2_U4_U6}. In particular we plot 
$j$ only over the period $-\pi<\Phi<\pi$, since all the quantities are 2$\pi$ periodic. The figure confirms the linearity of the persistent
currents with the magnetic flux for all the interaction strengths considered. The same is also true for other fillings for the 12 site ring (not 
presented here) away from half-filling. We also observe that the magnitude of persistent currents reduces with increasing $U/t$ for both ED 
and CLDFT and that the precise dependence of $j$ on $U/t$ is different for different fillings. This is in good agreement with previous 
calculations based on the BA technique \cite{BoBo}.

\begin{figure}[htb]
\begin{center}
\includegraphics[width=0.48\textwidth,angle=0.0]{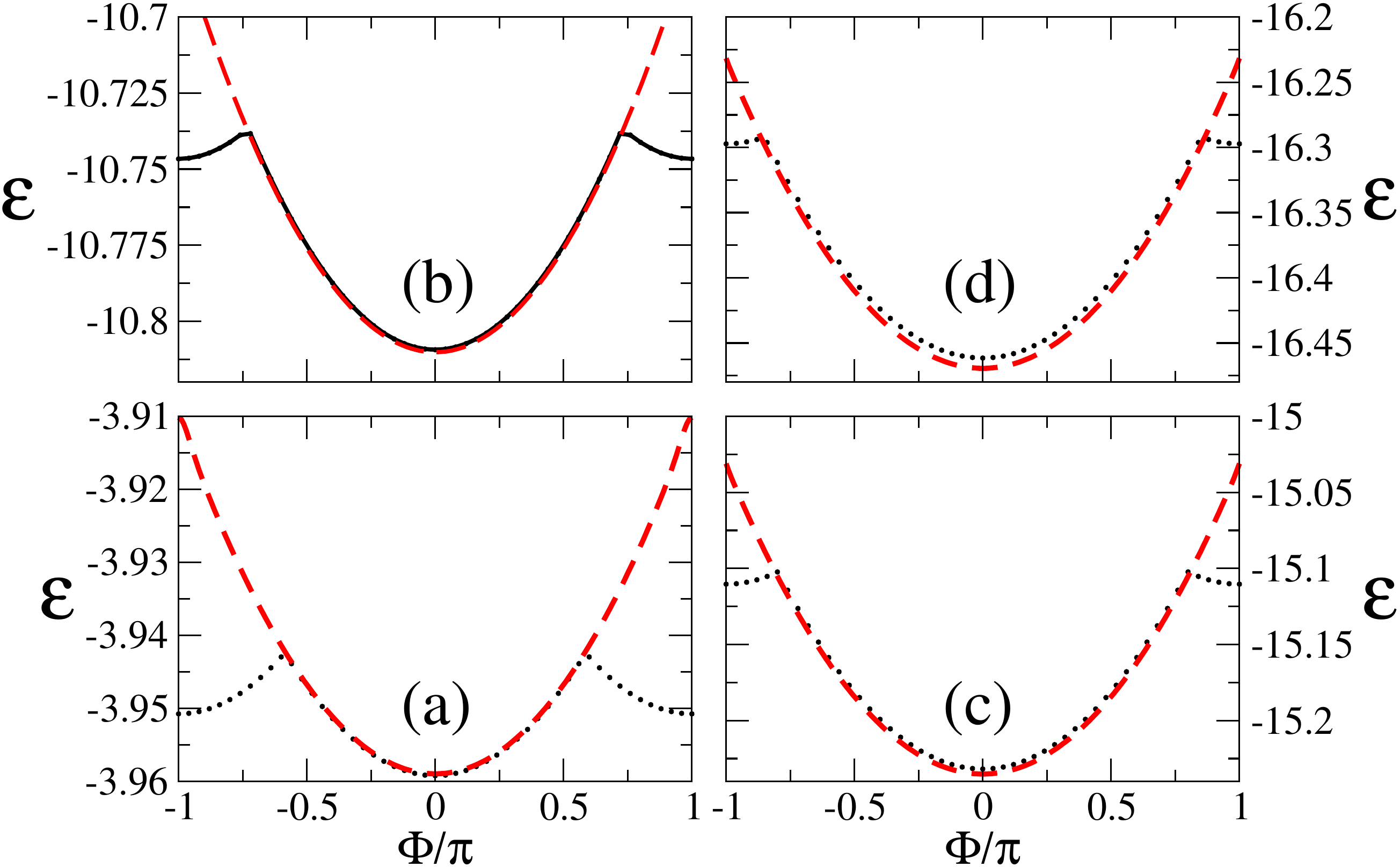}
\end{center}
\caption {(Color online) Ground state energy, $\mathcal{E}(\Phi)$, as a function of the magnetic flux, $\Phi$, calculated with both the
BA technique (dotted black line) and CLDFT (dashed red line). Calculations are carried out for $L=20$, $U/t=4$ and different numbers of 
electrons: (a) $N=2$ ($n=1/10$), (b) $N=6$ ($n=0.3$), (c) $N=10$ ($n=1/2$) and (d) $N=14$ ($n=0.7$).} \label{20_sites_2_6_10_14}
\end{figure}

ED is computationally demanding and cannot be performed beyond a certain system size. For this reason, in order to benchmark CLDFT for 
larger rings, we have calculated the ground state energy with the BA method. An example of these calculations is presented in figure 
\ref{20_sites_2_6_10_14}, where once again we show $\mathcal{E}(\Phi)$ for $L=20$, $U/t=4$ and different numbers of electrons. Also 
in this case the agreement between the BA results and those obtained with CLDFT is remarkably good as long as the ground state is a 
singlet. Interestingly we note that the agreement is better for low filling but it deteriorates as one approaches the half-filling case ($N=20$ 
in this case). This is somehow expected given the discontinuity of $\Lambda^\mathrm{xc}$ and of the derivative of $e^\mathrm{xc}$ at $n=1$
(see figure~\ref{exchange_lambda}), leading to the Mott transition. The presence of these discontinuities, although qualitatively correct, poses 
numerical problems and losses in accuracy.  

The final quantity we wish to consider is the charge stiffness or Drude weight, $D_\mathrm{c}$, defined as
\begin{equation}\label{chargestiff}
D_\mathrm{c}=\frac{L}{2}\frac{\partial^2\mathcal{E}(n,\Phi)}{\partial\Phi^2}|_{\Phi=0}\:.
\end{equation}
This is essentially the slope of the persistent current as a function of $\Phi$ calculated at $\Phi=0$ and defines the magnitude of the real part of the
optical conductivity in the long wave-length limit (see appendix for more details). $D_\mathrm{c}$ determines both qualitatively and quantitatively 
the transport properties of the ring. Importantly in the limit of large rings it exponentially vanishes for insulators, while it saturates to a finite value
for metals. Many studies have been devolved to calculating $D_\mathrm{c}$ for interacting systems. R\"{o}mer and Punnoose have studied 
$D_\mathrm{c}$ for finite Hubbard rings using an iterative BA technique \cite{Romer}. Eckern \textit{et. al.} explored the relation between 
$D_\mathrm{c}$ and the so-called phase sensitivity, $\Delta\mathcal{E}$,  for spinless fermions. $\Delta\mathcal{E}$ is the difference in the total 
energy calculated at $\Phi=0$ (periodic ground state) and that at $\Phi=\pi$ (antiperiodic ground state) \cite{Schenk1, Schmitteckert}. 
Recently a density matrix renormalization group algorithm has been developed to deal with complex Hamiltonian matrices and used to calculate 
$D_\mathrm{c}$ for spinless fermions \cite{Dias}.

\begin{figure}
\begin{center}
\includegraphics[width=0.40\textwidth,angle=0.0]{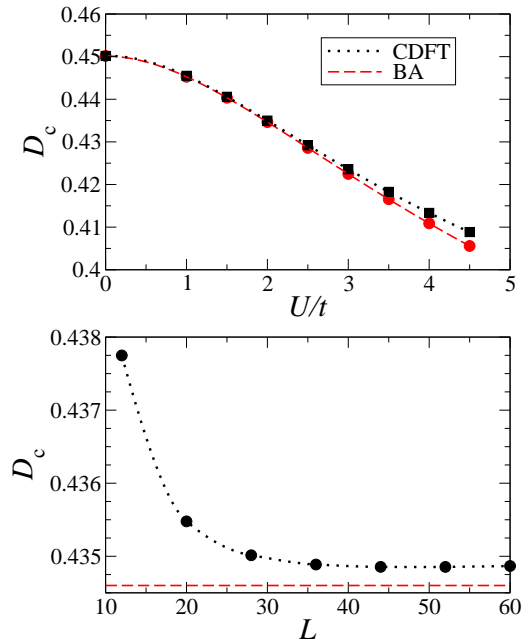}
\end{center}
\caption {(Color online) Drude coefficient $D_\mathrm{c}$ as a function of the interaction strength $U/t$ (top panel) and of the number of sites in the
ring, $L$ (bottom panel). All the calculations are for quarter filling and the results in the top panel are for a 60-site ring. In the figure we compare
CLDFT results (dotted black lines) with those obtained by the BA technique in the thermodynamic limit (dashed red lines). Calculations in the 
lower panel are for $U/t=2$.} 
\label{Dc_scheme}
\end{figure}

Since the agreement between CLDFT and ED is proved for small rings (the slopes of the persistent currents as a function of $\Phi$ calculated 
with CLDFT and ED are essentially identical in figure \ref{12_6persistentU0_U2_U4_U6}) we concentrate here on a larger system, namely a 
homogeneous 60 site ring at quarter filling. Our results for  the Drude weight as a function of $U/t$ are presented in figure \ref{Dc_scheme}. 
Again the CLDFT data are compared with those calculated with the BA in the thermodynamic limit ($L\rightarrow\infty$) and the agreement 
is rather satisfactory. We note that, as for the ground state energy, also for the Drude weight the CLDFT seems to perform less well as $U/t$ 
increases, i.e. as the interaction strength becomes large.  
Then in the lower panel of figure \ref{Dc_scheme} we illustrate the scaling properties of $D_\mathrm{c}$ as a function of the number of sites
in the ring, $L$ (we consider quarter filling and $U/t=2$). Clearly $D_\mathrm{c}$ does not vanish at any lengths demonstrating that the system 
remains metallic. Furthermore it approaches a constant value already for $L>40$. In the picture we also report the asymptotic value predicted 
by the BA in the thermodynamic limit $L\rightarrow\infty$ for this set of parameters. We find that the calculated CLDFT value is only 0.06\% larger
than the BA one, i.e. it is in quite remarkable good agreement. 

\subsection{Scaling properties}

\begin{figure}[htb]
\begin{center}
\includegraphics[width=0.48\textwidth,angle=0.0]{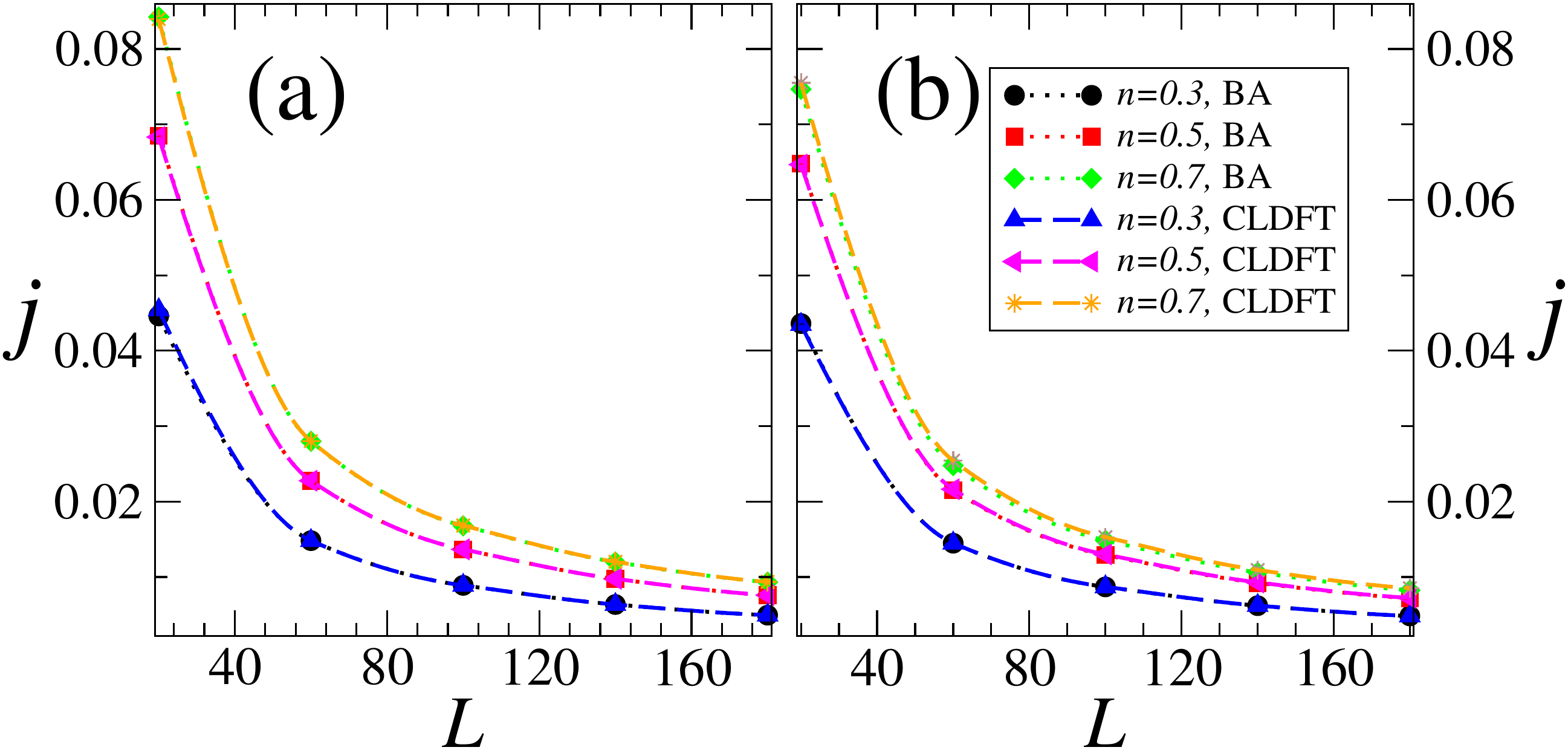}
\end{center}
\caption {(Color online) Persistent current, $j$, as a function of the number of site in the ring, $L$, and for different electron occupations, $n$:
(a) $U/t=2$, (b) $U/t=4$. Results are obtained with both the exact BA technique and CLDFT. In the figure the persistent currents are 
calculated at $\Phi=\pi/2$, i.e. $j=j(\pi/2)$} \label{different_sizes_U4}
\end{figure}

Next we take a more careful look at the scaling properties of the persistent currents and the Drude weights as a function of both the ring 
size and the interaction strength. It is well known that $j$ is strongly size dependent, since it originates from electron coherence across 
the entire ring \cite{Dias}. For a perfect metal one expect $j$ to scale as $1/L$ \cite{Bouzerar}. In Figure \ref{different_sizes_U4} the 
value of the persistent currents as a function of the ring size are presented for different electron fillings and for the two representative 
interaction strengths of $U/t=2$ (a) and $U/t=4$ (b). Calculations are performed with both the exact BA and CLDFT. As a matter
of convention we calculate the persistent currents at $\Phi=\pi/2$.

In general we find a monotonic reduction of the persistent current with $L$ and an overall excellent agreement between the BA and the 
CLDFT results over the entire range of lengths, occupations and interaction strengths investigated. 
A non-linear fit of all the curves of figure~\ref{different_sizes_U4} returns us an almost perfect $1/L$ dependence of $j$ with no 
appreciable deviations at any $n$ or $U/t$. This indicates a full metallic response of the rings in the region of parameters investigated,
thus confirming previous results obtained with the BA approach \cite{BoBo}.

\begin{figure}[htb]
\begin{center}
\includegraphics[width=0.5\textwidth,angle=0.0]{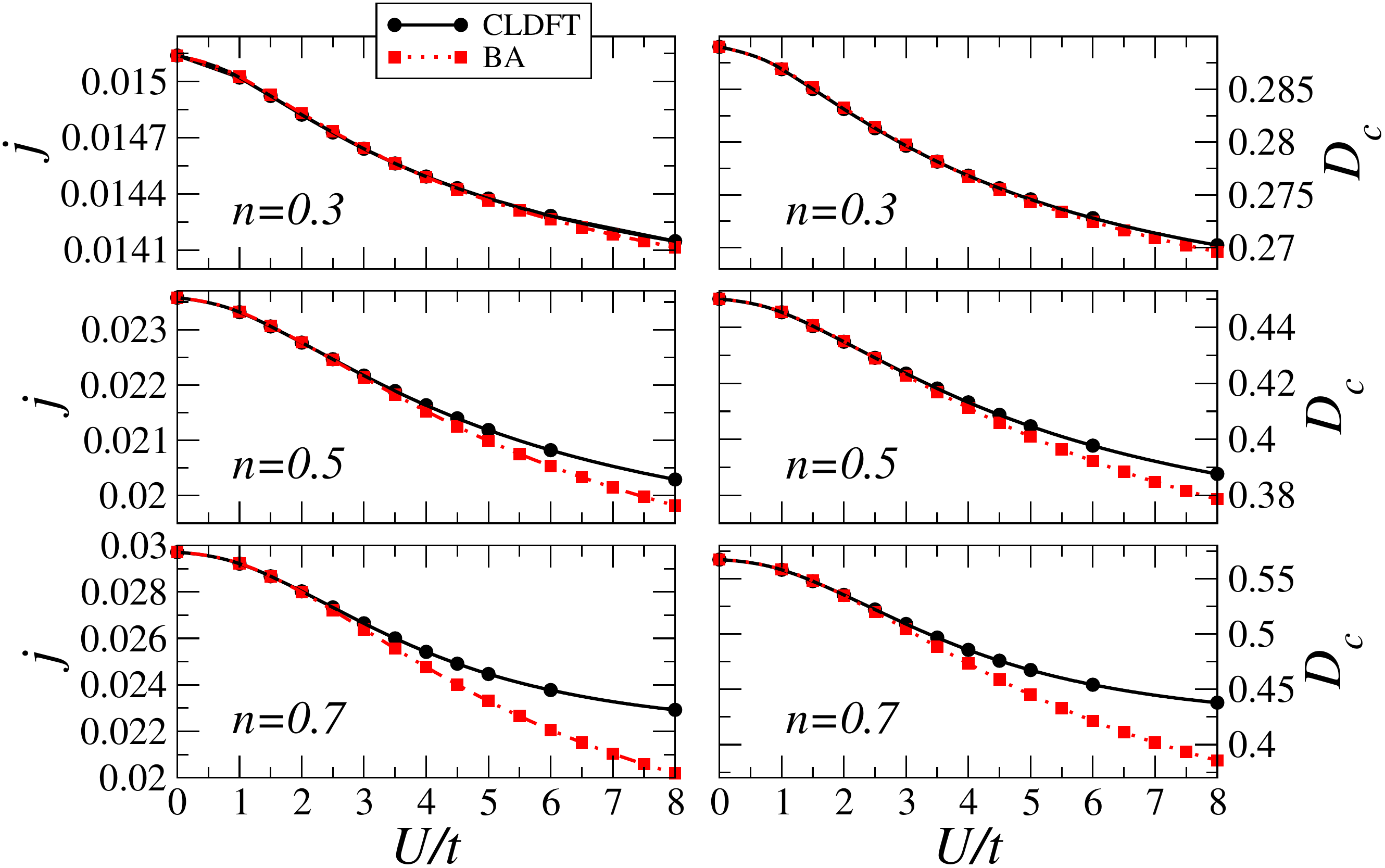}
\end{center}
\caption {(Color online) Persistent current, $j$, and Drude weight, $D_\mathrm{c}$ as a function of interaction strength $U/t$ for a
60 site ring at different fillings. Results are obtained with both the exact BA technique and CLDFT.} \label{PCandDW}
\end{figure}

Then we look at the dependance of $j$ and $D_\mathrm{c}$ on the interaction strength. In this case we consider a 60 site ring and 
three different different electron fillings. In general for small fluxes one expects $j=2D_\mathrm{c}\Phi$ and our numerical results of 
Fig.~\ref{PCandDW} demonstrates that this is approximately correct also for our definition of persistent currents [$j=j(\Phi=\pi/2)$] 
over the entire $U/t$ range investigated. 
We find that both $j$ and $D_\mathrm{c}$ monotonically decrease as a function of the interaction strength, essentially meaning that 
the predicted long-wavelength optical conductivity is reduced as the electron repulsion gets larger. 

Also in this case the agreement between the BA and the CLDFT results is substantially good, although significant deviations appear
in the limit of large $U/t$ and electron filling approaching half-filling. This again corresponds to a region of the parameter space where 
the XC potential approaches the derivative discontinuity. 

It was numerically demonstrated in the past \cite{BoBo} that the persistent current (and so the Drude weight) at half-filling follows the scaling 
relation $j\sim\mathrm{e}^{-U^2/\xi}$, with $\xi\sim1$. However, to the best of our knowledge, no scaling relation was ever provided in the 
metallic case. We have then carried out a fitting analysis (the fit is limited to values of $j$ and $D_\mathrm{c}$ for $U/t>2$) and found that our 
data can be well represented by the scaling lows
\begin{equation}\label{esl}
j=j_0(U/t)^{-\beta},\;\;\;\;\;\;\;D_\mathrm{c}=D_0(U/t)^{-\gamma}\:.
\end{equation}
In general and as expected we find $\beta=\gamma$ and a quite significant dependence of the exponents on the filling. In particular
table~\ref{thetable} summarizes our results and demonstrates that the decay rate of both the persistent currents and the Drude weights
increases as the filling approaches half-filling. Furthermore the table also quantifies the differences between the BA and the CLDFT
solutions, whose exponents increasingly differ from each other as the electron filling gets closer to $n=1$ (for $n=0.7$ we find 
$\beta^\mathrm{BA}\sim2\beta^\mathrm{CLDFT}$).
 \begin{table}[h]
\begin{tabular}{ccc} \hline\hline
$n$ & $\beta^\mathrm{CLDFT}$ & $\beta^\mathrm{BA}$ \\ \hline
0.3 & 0.036 & 0.036  \\ 
0.5 & 0.085 & 0.104  \\ 
0.7 & 0.151 & 0.246  \\ \hline\hline
\end{tabular}\caption{\label{thetable} Exponents for the empirical scaling laws of equation (\ref{esl}) as fitted from the data of figure
\ref{PCandDW}.}
\end{table}

Finally, by combining all the results of this section we can propose a scaling law for both the persistent currents and the
Drude weights, valid in the metallic limit of the Hubbard model, i.e. away from half-filling. This reads
\begin{equation}
j=\frac{j_0(n)}{L}\left(\frac{U}{t}\right)^{-\beta(n)}\:,
\end{equation}
where both the constant $j_0$ and the exponent $\beta$ are function of the electron filling $n$. Note that an identical equation
holds for $D_\mathrm{c}$.

\subsection{Scattering to a single impurity}

Having established the success of the BALDA to CLDFT for the homogeneous case we now move to a more stringent test for the theory, namely
the case of a ring penetrated by a magnetic flux in the presence of a single impurity. Such a problem has already received considerable 
attention in the past \cite{Viefer2, Koskinen, Cheung}. Note that, as in {\it ab initio} DFT, this is a situation different from the reference system used 
to construct the BALDA (since it deals with a non homogeneous system) and therefore one might expect a more pronounced disagreement with 
exact results. As the BA equations are integrable only for the homogeneous case we now benchmark our CLDFT results with those obtained by 
ED. This however limits our analysis to small rings. 

\begin{figure}[htb]
\begin{center}
\includegraphics[width=0.48\textwidth,angle=0.0]{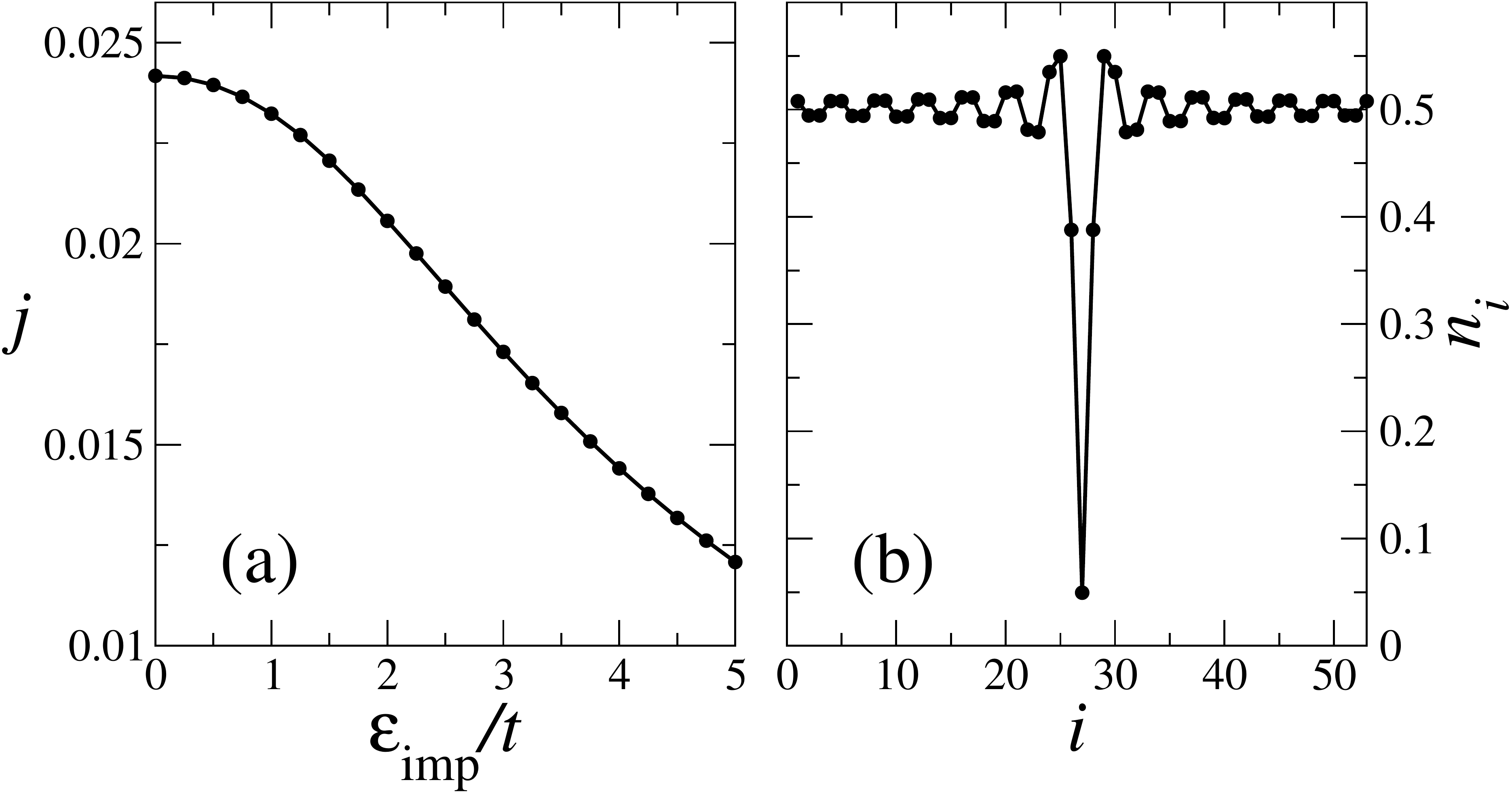}
\end{center}
\caption {(a) Persistent current, $j$, as a function of single impurity strength, $\varepsilon_\mathrm{imp}$, obtained from the CLDFT for $L=53$, $N=26$, 
$U/t=4$ and $\Phi=\frac{\pi}{2}$. In (b) we show a typical site density profile for a positive single impurity site potential.} \label{53_26_persistent_vs_e1}
\end{figure}

The single impurity in the ring is described by simply adding to the Hamiltonian of equation (\ref{Hint}) the term
\begin{equation}\label{impurity_ham}
\begin{aligned}
\hat{H}_\mathrm{imp}=&\varepsilon_\mathrm{imp}\hat{n}_{i}\:,
\end{aligned}
\end{equation}
where $\varepsilon_\mathrm{imp}$ is the modification to the on-site energy at the impurity site $i$. The inclusion of an impurity produces in general
electron backscattering so that we expect the persistent currents to get reduced. In figure \ref{53_26_persistent_vs_e1} we present the general 
transport features for this inhomogeneous system. Calculations have been carried out with CLDFT for a ring comprising 53 sites and $N=26$, 
$U/t=4$. Again the persistent currents are calculated at $\Phi=\pi/2$.

Panel (a) shows $j$ as a function of the impurity on-site energy. As expected from standard scattering theory the current is reduced as 
$\varepsilon_\mathrm{imp}$ increases, thus creating a potential barrier. The electron density profile for this situation is presented in panel (b), where
one can clearly observe an electron depletion at the impurity site and Friedel's oscillations around it. 

\begin{figure}[htb]
\begin{center}
\includegraphics[width=0.48\textwidth,angle=0.0]{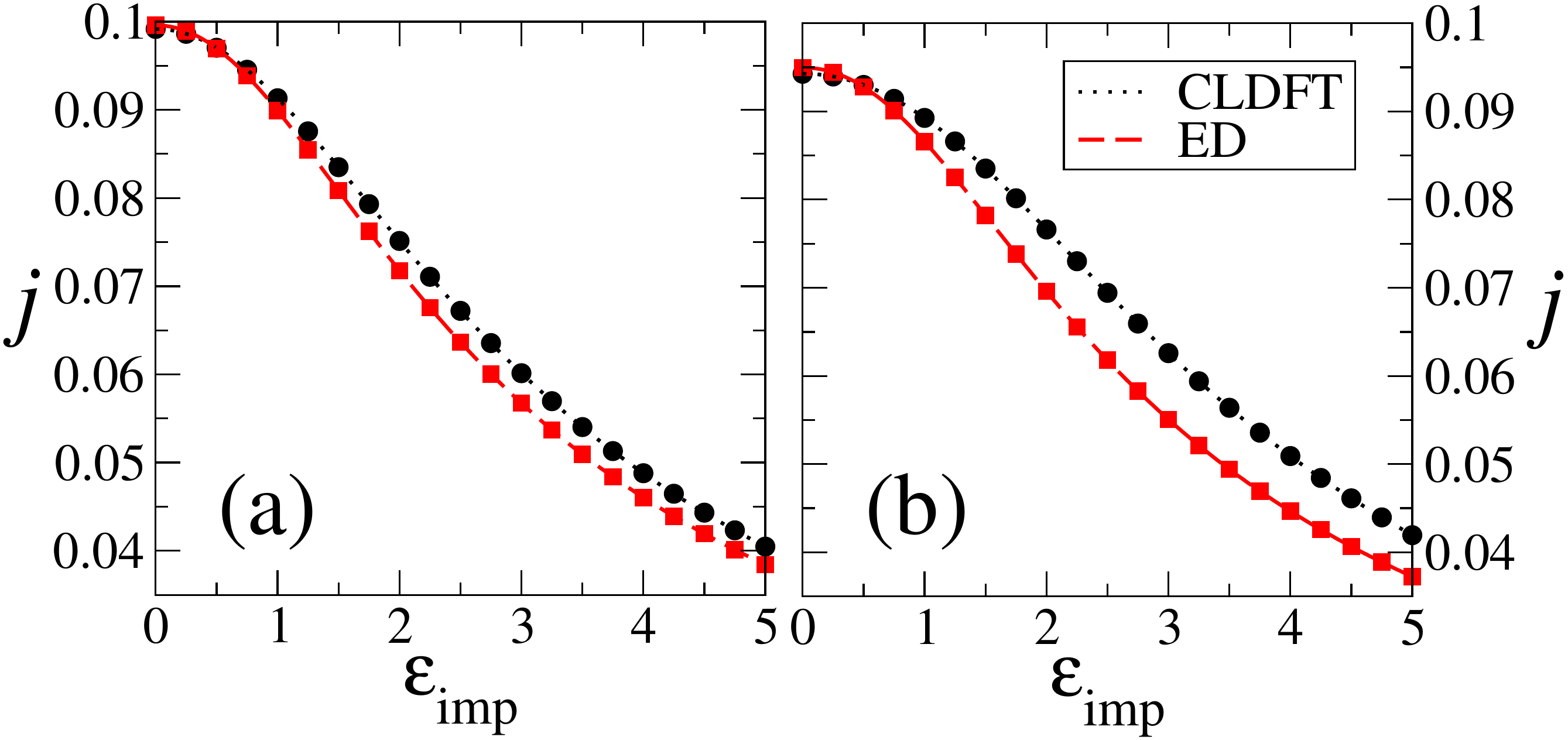}
\end{center}
\caption {(Color on line) Comparison between the persistent currents calculated with CLDFT (black symbols and dotted line) and by ED 
(red symbols and dashed line) for a 13 site ring and $N=6$. The $j$'s are obtained at $\Phi=\pi/2$ for two different values of the interaction 
strength, namely $U/t=2$ (a) and $U/t=4$ (b).} \label{BAvsCLDFT}
\end{figure}

A quantitative assessment of our CLDFT results is provided in Fig.~\ref{BAvsCLDFT} where they are compared with those obtained by exact 
diagonalization for a 13 site ring close to quarter filling ($N=6$). In particular we present $j$ as a function of the impurity potential, 
$\varepsilon_\mathrm{imp}$, for both $U/t=2$ and $U/t=4$. In general we find a rather satisfactory agreement between CLDFT and the
exact results in particular for small $\varepsilon_\mathrm{imp}$ and $U/t$. As the electron scattering becomes more significant deviations
appear and the quantitative agreement is less good. Importantly we notice that the ED results systematically provide a persistent current 
lower than that calculated with CLDFT, at least for the values of electron filling investigated here. This seems to be a consistent trend also present 
for the homogeneous case (see figure~\ref{PCandDW}), although the deviations in that case are less pronounced (for the same electron filling and
interaction strength). Therefore we tentatively conclude that most of the errors in the impurity problem have to be attributed to the errors already 
present in the homogeneous case. We then expect that CLDFT provides a good platform for investigating scattering problems at only minor 
computational costs. As such CLDFT appears as the ideal tool for investigating the interplay between electron-electron interaction and 
disorder in low dimensional structures. 

\section{Conclusion}
In this work we have presented an extension of the BALDA for the one-dimensional Hubbard problem on a ring to CLDFT. We have then investigated
the response of interacting rings to an external flux both in the homogeneous and inhomogeneous case, and we have compared our results with
those obtained by numerically exact techniques. Our analysis has been confined to the metallic limit, i.e. away from half-filling, where the Hubbard model has
a metal to insulator transition. In general we have found that CLDFT performs rather well in calculating both the persistent currents and the 
Drude weights in the homogeneous case. Furthermore a similar level of accuracy is transferred to the impurity problem. With these results in hands 
we propose to use CLDFT in the study of AB rings where the combined effect of electron-electron interaction and disorder can be addressed 
for large rings, so that a numerical evaluation of the various scaling laws proposed in the past can be accurately carried out.

\section{Acknowledgements}
We thank N. Baadji, I. Rungger and V. L. Campo for useful discussions. This work is supported by Science Foundation of Ireland under the 
grant SFI05/RFP/PHY0062 and 07/IN.1/I945. Computational resources have been provided by the HEA IITAC project managed by the Trinity 
Center for High Performance Computing and by ICHEC.

\section{APPENDIX: Local Density Approximation for the CLDFT}
We use the BA solution for the homogeneous part of the $\hat{H}_U^\Phi$ [equation~(\ref{Hint})] to estimate the XC energy. Then the local 
approximation is taken,
\begin{equation}
\mathcal{E}^\mathrm{xc}_\mathrm{LDA}[n_l,j_l]=\sum_le^\mathrm{xc}[n_l,j_l]\:.
\end{equation}
Here $e^\mathrm{xc}(=\frac{\mathcal{E}^\mathrm{xc}[n,j]}{L})$ is the XC energy per site for the homogeneous system, which is provided
in equation (\ref{exc3}). The first term of the equation (\ref{exc3}) can be calculated exactly using the BA procedures \cite{Shastry} to obtain 
the ground state energy as a function of $n$ and $\Phi$. Then the phase variable $\Phi$ can be eliminated from the ground state energy 
to contain the current via
\begin{equation}
j=\frac{\partial \mathcal{E}(n,\Phi)}{\partial \Phi}\:. 
\end{equation}

The complete flux dependence of the ground state energy for the Mott insulator phase ($n=1$) in the thermodynamic limit has been shown 
to be \cite{Stafford}
\begin{equation}
\mathcal{E}(n,\Phi)-\mathcal{E}(n,0)=\frac{2D_c(n)}{L}(1-\cos\Phi)\:, 
\end{equation}
while away from half filling and $L\rightarrow\infty$ this is
\begin{equation}\label{fluxdependence2}
\mathcal{E}(n,\Phi)-\mathcal{E}(n,0)=\frac{D_\mathrm{c}(n)}{L}\Phi^2\:. 
\end{equation}
Here $D_\mathrm{c}(n)$ is the charge stiffness (Drude weight) defined as 
\begin{equation}\label{chargestiff}
D_\mathrm{c}=\left.\frac{L}{2}\frac{\partial^2\mathcal{E}(n,\Phi)}{\partial\Phi^2}\right|_{\Phi=0}. 
\end{equation}
In physical terms the Drude weight $D_\mathrm{c}$ is the real part of the optical conductivity $\sigma_1(w)$ in the long wavelength 
limit \cite{Stafford},
\begin{equation}
 \sigma_1(w)=2\pi D_\mathrm{c}\delta (w)+\sigma_1^{\text{reg}}(w)\:,
\end{equation}
where we took $\hbar=e=c=1$. If we denote $\mathcal{E}^\mathrm{BA}(n_\mathrm{BA},\Phi_\mathrm{BA})$ and $\mathcal{E}^0(n_{0},\Phi_{0})$ 
respectively as the ground state energy for the interacting system [first term in equation (\ref{exc3})] and for the non-interacting one [second term 
in equation (\ref{exc3})], away from half-filling we will write
\begin{equation}\label{fluxdependence3}
\begin{aligned}
\mathcal{E}^\mathrm{BA}(n_\mathrm{BA},\Phi_\mathrm{BA})=&\mathcal{E}^\mathrm{BA}(n_\mathrm{BA},0)+\frac{D_\mathrm{c}^\mathrm{BA}(n_\mathrm{BA})}{L}\Phi^{2}_\mathrm{BA}\:,\\
\mathcal{E}^0(n_{0},\Phi_{0})=&\mathcal{E}^0(n_{0},0)+\frac{D_c^0(n_{0})}{L}\Phi^{2}_0\:,
\end{aligned}
\end{equation}
and
\begin{equation}\label{jdependence3}
\begin{aligned}
j^\mathrm{BA}(n_\mathrm{BA},\Phi_\mathrm{BA})=&2\frac{D_\mathrm{c}^\mathrm{BA}(n_\mathrm{BA})}{L}\Phi_\mathrm{BA}\:,\\
j^0(n_{0},\Phi_{0})=&2\frac{D_c^0(n_{0})}{L}\Phi_0\:.
\end{aligned}
\end{equation}
The fundamental requirement of the KS mapping is that $n_\mathrm{BA}=n_{0}=n$ and $j^\mathrm{BA}=j^{0}=j$ while we note that 
$\Phi_\mathrm{BA}=\Phi$ and $\Phi_{0}=\Phi^s$ in equation (\ref{exc3}). By substituting equation (\ref{fluxdependence3}) and 
the expressions for $\Phi^s$ and $\Phi$ obtained from equation (\ref{jdependence3}) into equation (\ref{exc3}) one obtains
\begin{equation}\label{exc4}
\mathcal{E}^\mathrm{xc}(n,j)=\mathcal{E}^\mathrm{BA}(n,0)-\mathcal{E}^0(n,0)-\mathcal{E}^\mathrm{H}(n)+\frac{L}{2}\Lambda^\mathrm{xc}(n)j^2,
\end{equation}
where
\begin{equation}\label{lambda}
\Lambda^\mathrm{xc}(n)=\frac{1}{2}\left [\frac{1}{D_\mathrm{c}^0(n)}-\frac{1}{D_\mathrm{c}^\mathrm{BA}(n)}\right ].
\end{equation}
Here $D_\mathrm{c}^0(n)$ is the non-interacting charge stiffness defined as
\begin{equation}
 D_\mathrm{c}^0(n)=\frac{2t}{\pi}\sin\left (\frac{n\pi}{2}\right )
\end{equation}
for $L\rightarrow\infty$. $D_\mathrm{c}^\mathrm{BA}(n)$ can then be obtained in the thermodynamic limit by using \cite{Kawakami}
\begin{equation}\label{Dccalc}
 D_\mathrm{c}^\mathrm{BA}(n)=\frac{1}{2\pi}[\xi_\mathrm{c}(Q)]^2v_\mathrm{c}
\end{equation}
where $\xi_\mathrm{c}$ is an element of the dressed charge matrix, which is used to describe the scattering between the quasi-particles and 
$v_\mathrm{c}$ is velocity of the charge excitation. Therefore one finally obtains
\begin{equation}\label{exc5}
e^\mathrm{xc}(n,j)=e^\mathrm{xc}(n,0)+\frac{1}{2}\Lambda^\mathrm{xc}(n)j^2,
\end{equation}
so that
\begin{equation}\label{vxc}
v^\mathrm{xc}_\mathrm{BALDA}(n_l,j_l)=\left.\frac{\partial e^\mathrm{xc}(n,j)}{\partial n}\right|_{n\rightarrow n_l,j\rightarrow j_l},
\end{equation}
and
\begin{equation}\label{phixc}
\Phi^\mathrm{xc}_\mathrm{BALDA}(n_l,j_l)=\left.\frac{\partial e^\mathrm{xc}(n,j)}{\partial j}=\Lambda^\mathrm{xc}(n)j\right|_{n\rightarrow n_l,j\rightarrow j_l}.
\end{equation}

\end{document}